\documentclass[twocolumn,prl,superscriptaddress,showpacs]{revtex4}

\usepackage{graphicx}
\usepackage{times}
\usepackage{amsmath,amssymb}
\usepackage{amsbsy}

\newcommand{\degree}{\ensuremath{^\circ}}

\newcommand{\ket}[1]{\left|{#1}\right\rangle}

\begin{document}

\title{Orbital ordering in e$_{\bf g}$ orbital systems: Ground states and thermodynamics of the 120\degree\ model}

\author{Andre van Rynbach}
\affiliation{Physics Department, University of California, Santa Barbara, California 93106}
\author{Synge Todo}
\affiliation{Department of Applied Physics, University of Tokyo, Tokyo 113-8656, Japan}
\affiliation{CREST, Japan Science and Technology Agency, Kawaguchi 332-0012, Japan}
\author{Simon Trebst}
\affiliation{Microsoft Research, Station Q,
University of California, Santa Barbara, CA 93106} 

\date{\today}

\begin{abstract}
Orbital degrees of freedom shape many of the properties of a wide class of Mott insulating, transition metal oxides with partially filled $3d$-shells. Here we study orbital ordering transitions in systems where a single electron occupies the $e_g$ orbital doublet and the spatially highly anisotropic orbital interactions can be captured by an orbital-only model, 
often called the 120\degree\ model.
Our analysis of both the classical and quantum limits of this model in an extended parameter space shows
that the 120\degree\ model is in close proximity to several $T=0$ 
phase transitions and various competing ordered phases. 
We characterize the orbital order of these nearby phases and their associated thermal phase transitions 
by extensive numerical simulations and perturbative arguments.
\end{abstract}

\pacs{71.20.Be, 05.70.Fh, 75.25.Dk}
\maketitle


Mott insulating transition metal oxides with partially filled $3d$-shells -- such as the manganites  -- exhibit rich phase diagrams with many competing orders, indicating a non-trivial interplay of spin, charge, and orbital degrees of freedom \cite{ScienceNagaosa}.
A prominent example of a material exhibiting {\sl orbital order} is the extensively studied LaMnO$_3$ \cite{LaMnO3}.
The crystal field in this perovskite material splits the five $d$-orbitals into three t$_{2g}$ orbitals occupied by three electrons,
and an $e_g$ doublet sharing a single electron. 
This partially filled $e_g$ doublet then gives rise to an additional orbital degree of freedom indicating which of the two orbitals 
is occupied. The exchanges between these orbital degrees of freedom -- arising from Jahn-Teller distortions or Kugel-Khomskii type superexchange --
are oftentimes described by orbital-only models which neglect the spin degrees of freedom. The latter is justified in situations where the energy scales
of spin and orbital interactions are well separated, i.e. orbital interactions correspond to temperature scales where the spins are still largely disordered
or in a situation where the spins are effectively frozen out (e.g. by a magnetic field).
Expressing the  $e_g$ orbital degree of freedom by a two-component pseudospin 
$\mathbf{T} = (T^z, T^x)$, where $T^z= \pm1$ correspond to occupation of the $\ket{3z^2-r^2}$ and $\ket{x^2-y^2}$ orbitals,
the highly anisotropic interactions between them are captured by the so-called 120\degree\ model \cite{120model}
on a cubic lattice 
\begin{eqnarray}
\label{eq:H120}
 H_{120} = &-\sum_{i,\gamma = x,y} \frac{1}{4} \left[ J_z T_i^z T_{i+\gamma}^z + 3 J_x T_i^x T_{i+\gamma}^x \phantom{\sqrt{3}J_{\rm mix}} \right.\\ 
                  & \left. \pm \sqrt{3}J_{\rm mix}(T_i^z T_{i+\gamma}^x+T_i^x T_{i+\gamma}^z)\right] - \sum_i J_z T_i^zT_{i+z}^z\,, \nonumber
\end{eqnarray}
where the $\pm$ sign for the `mixing' term enters for coupling along the $x$ and $y$ directions, respectively \cite{FootnoteSign}.
If the orbital exchange is primarily mediated through Jahn-Teller distortions, this model is commonly considered in its classical limit, 
where the pseudospins $\mathbf{T}$ are $O(2)$ spins. 
If, on the other hand, the orbital exchange arises primarily from a Kugel-Khomskii type superexchange \cite{KK}, this model should be considered in
its quantum limit. In the latter case, the pseudospins $\mathbf{T}$ are identified with $SU(2)$ spins, 
i.e.~their components  become Pauli matrices $T^{x,z} = \frac{1}{2}\sigma^{x,z}$. 
The above 120\degree\ model has typically been studied at equal coupling $J_x = J_z = J_{\rm mix}$, for which it exhibits 
an enhanced rotational symmetry where the symmetry of the cubic lattice under permutations of the $x$, $y$, and $z$ axes is 
reflected in a three-fold symmetry in the $(T^z, T^x)$ plane. 
This becomes apparent when rewriting \eqref{eq:H120} as 
$
 H_{120} = -J \sum_{i,\gamma = x,y,z} (\boldsymbol\tau_i \cdot \mathbf{e}^\gamma)(\boldsymbol\tau_{i+\gamma}\cdot\mathbf{e}^\gamma) \,,
$
where the $\mathbf{e}^\gamma$ are unit vectors in the $x,y,z$-directions and the $\boldsymbol\tau_i$ are defined as three-component vectors 
${\boldsymbol\tau_i} = \left( [T^z_i + \sqrt{3}T^x_i]/2,  [T^z_i - \sqrt{3}T^x_i]/2, T^z_i \right)$.

While the presence of this enhanced rotational symmetry for equal coupling 
has greatly benefitted the understanding of the classical model
and has led to a rigorous description of its highly degenerate ground-state manifold \cite{Biskup05}, it has remained elusive to identify the ground states 
of the quantum model solely based on symmetry arguments.
In this manuscript, we will take a broader perspective and study the above 120\degree\ model away from this symmetric point and
explore ground states and thermodynamic properties  in an extended two-dimensional parameter space $(J_x/J_z, J_{\rm mix}/J_z)$,
which in experiments should be accessible by changing pressure or adding a small electric field. 
Our approach reveals that the original 120\degree\ model is in close proximity to several $T=0$ phase transitions and various competing ordered 
phases.
Combining extensive numerical simulations with analytical arguments we describe the orbital order in these phases for the classical and quantum limits of this extended 120\degree\ model, as well as thermal phase transitions associated with these phases and $T=0$ phase transitions between them.


\paragraph{The classical model.--}

\begin{figure}[t]
\begin{center}
  \includegraphics[width=\columnwidth]{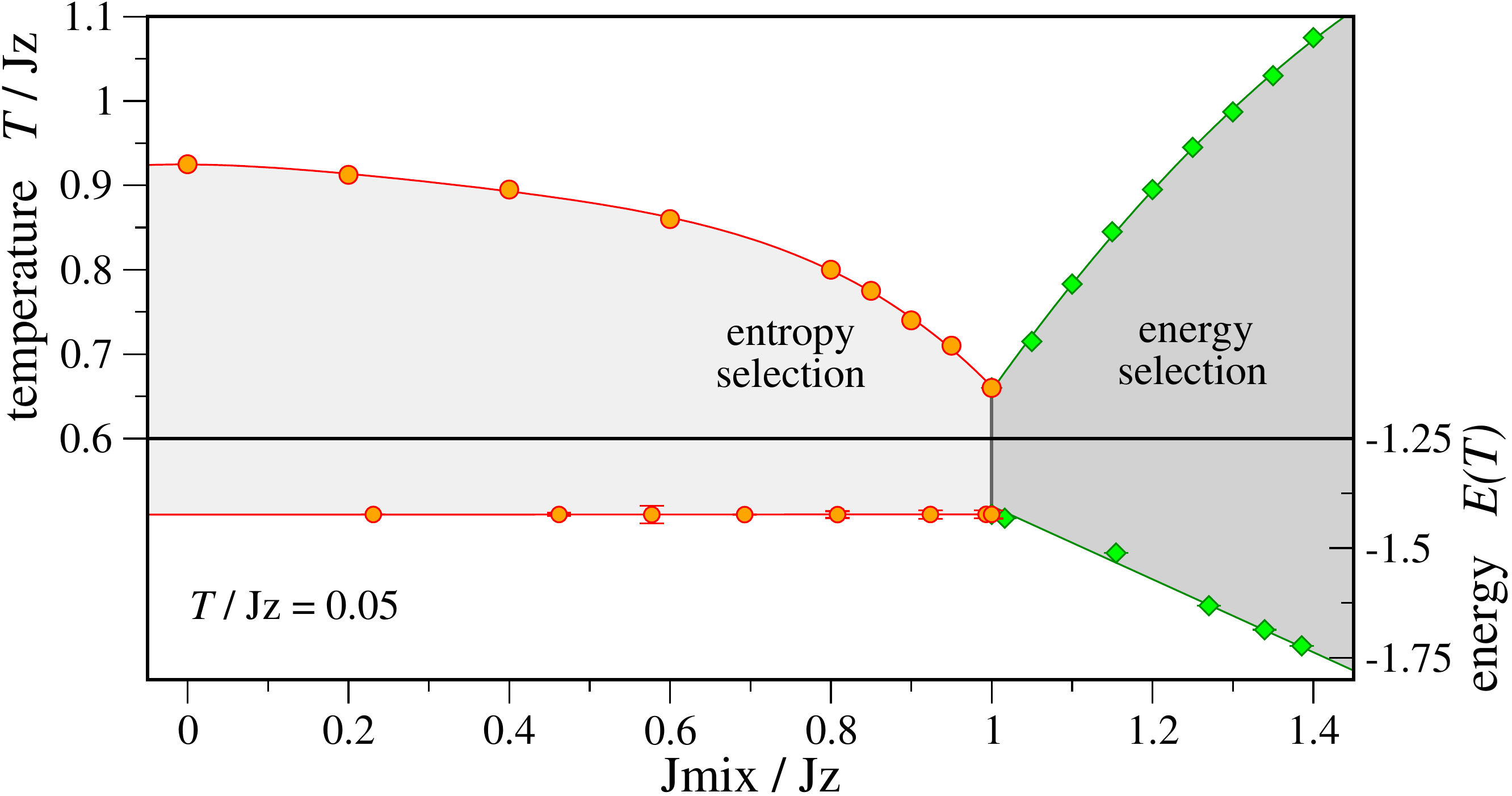}
\end{center}
\caption{
   Variation of the transition temperature (upper panel) and ground-state energy (lower panel)
   in the classical model for varying $J_{\rm mix}$.
   The 120\degree\ model corresponds to $J_{\rm mix} = J_z$.
     }
\label{Fig:Tc_Jmix}
\end{figure}

For the classical 120\degree\ model with rotational symmetry, e.g. $ J_x = J_z = J_{\rm mix}$,
it has long been appreciated that this model exhibits an infinite, but sub-extensive ground-state degeneracy \cite{Biskup05}, 
which is split at low temperatures by an order-by-disorder mechanism stabilizing six ordered states 
\cite{Europhys,Biskup05}.
Before turning to the question of how these characteristic features change when exploring the model in
the extended parameter space, we will briefly recount their origin in the symmetric model.
To this end, we label a general state in the orbital subspace by an angle 
  $\ket{\theta} = \cos(\theta/2) \ket{3z^2-r^2} + \sin(\theta/2) \ket{x^2-y^2}$,
which for the classical model simply describes the orientation of the $O(2)$ pseudospin vector.
\begin{figure}[b]
\begin{center}
  \includegraphics[width=\columnwidth]{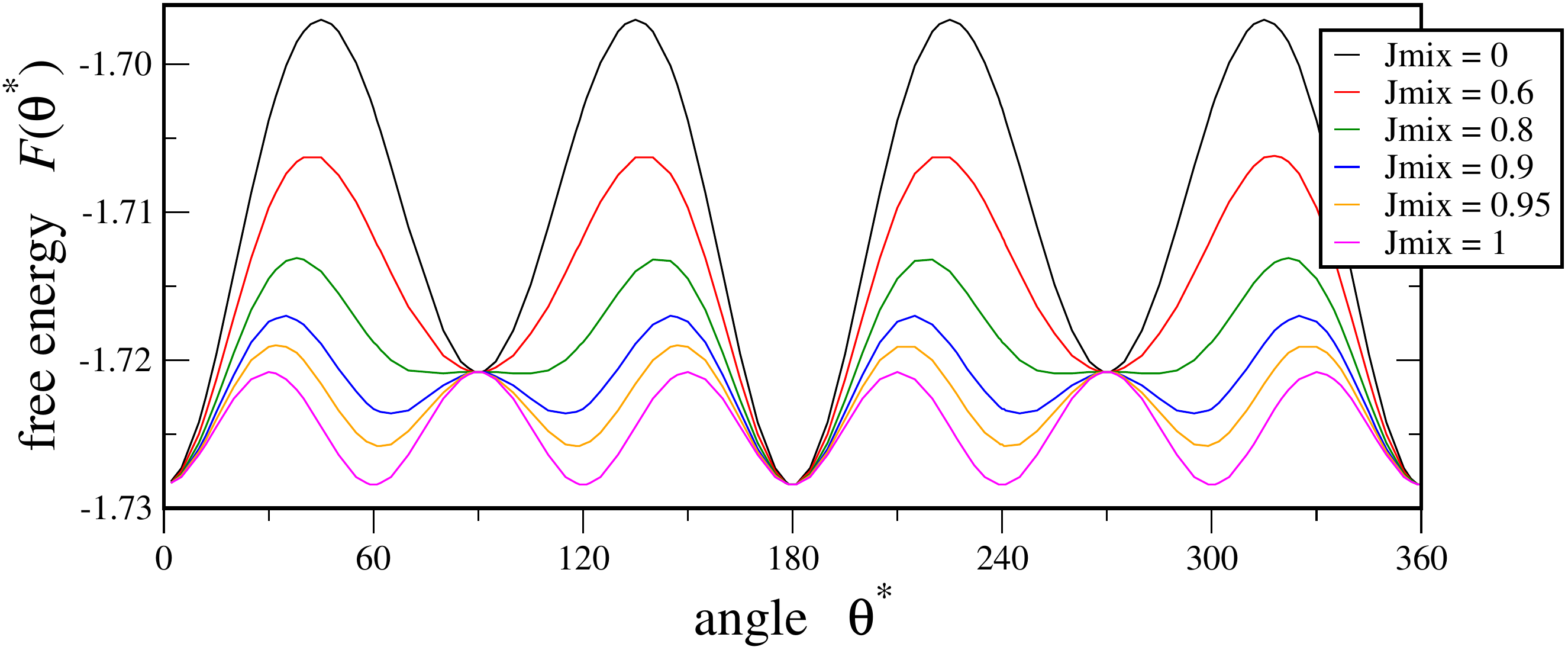}
\end{center}
\caption{
   Entropic selection of low-temperature states in the classical model: 
   The free energy $F(\theta^*)$ obtained from a spin-wave analysis of \eqref{eq:H120} 
   as a function of $J_{\rm mix}/J_z$ for fixed $J_x = J_z$.
  }
\label{Fig:F_Jmix}
\end{figure}
To identify the degenerate manifold of ground states, we first observe that any polarized
state with all pseudospins being aligned along some angle $\theta^*$ is a ground state of Hamiltonian \eqref{eq:H120}.
Starting from any such state, further ground states can be found~\cite{Europhys,Biskup05} by reflecting all orbitals 
in the $xy$ plane about a line at $0\degree$, the $xz$ plane about $120\degree$, or the $yz$ plane about $240\degree$.
Remarkably, we find that this ground-state degeneracy remains (partially) unscathed for an extended parameter regime when moving away from the 
symmetric model on a line described by $0 \leq J_{\rm mix} / J_z \leq 1$ and $J_x = J_z$.
What distinguishes states along this line in parameter space, however, is their instability to thermal fluctuations and the entropic 
selection of low-temperature states. To discuss this order-by-disorder mechanism we calculate the free energy of the low-temperature
states by considering a spin-wave approximation of \eqref{eq:H120} and expanding to second order in small fluctuations 
$\delta\theta_i = \theta_i - \theta^*$ about an orbitally ordered state with $\theta_i=\theta^*$ at each site. 
The resulting free energy $F(\theta^*)$ is plotted in Fig.~\ref{Fig:F_Jmix} as a function of $J_{\rm mix}/J_z$. 
For $J_{\rm mix} \lesssim 0.8~J_z$, the ground-state manifold is lifted and four low-temperature states are entropically favored with their free energy 
being minimized at angles $\theta^* = 0\degree,90\degree,180\degree,270\degree$. These four states correspond to orbitally ordered states
in orbital configurations given by 
$\ket{0\degree} = \ket{3z^2 - r^2}$,
$\ket{180\degree} = \ket{x^2 - y^2}$, and
$\left( \ket{3z^2-r^2} \pm \ket{x^2 -y^2} \right) / \sqrt{2}$ for the $90\degree$ and $270\degree$ states, respectively.
For $J_{\rm mix} \gtrsim 0.8~J_z$ the $90\degree$-minima in the free-energy curves bifurcate and in total form six minima, 
all of which become exactly equal only for the symmetric model $J_{\rm mix}=J_z$, where the minima are located precisely at angles of 
$\theta^* = 0\degree,60\degree,120\degree,\ldots,300\degree$.  
If we further enlarge $J_{\rm mix}$ beyond $J_z$, this order-by-disorder phenomenon disappears and we instead find that the ground 
and low-temperature states are {\sl energetically} selected, with the $xz$ or $yz$ planes ordering in alternating orientations 
of $\theta_1 $ and $\theta_1 + 180\degree$ where $\theta_1$ continuously changes from $\theta_1 \approx 30 \degree$ to $\theta_1 \approx 45 \degree$ with increasing $J_{\rm mix}$. 
This transition between entropic and energetic selection occurs
exactly at the symmetric point $J_x = J_z = J_{\rm mix}$ of the 120\degree\ model and is accompanied by a first-order phase transition at
zero temperature, which becomes apparent in a level-crossing of ground-state energies shown in 
Fig.~\ref{Fig:Tc_Jmix}.

\begin{figure}
\begin{center}
  \includegraphics[width=\columnwidth]{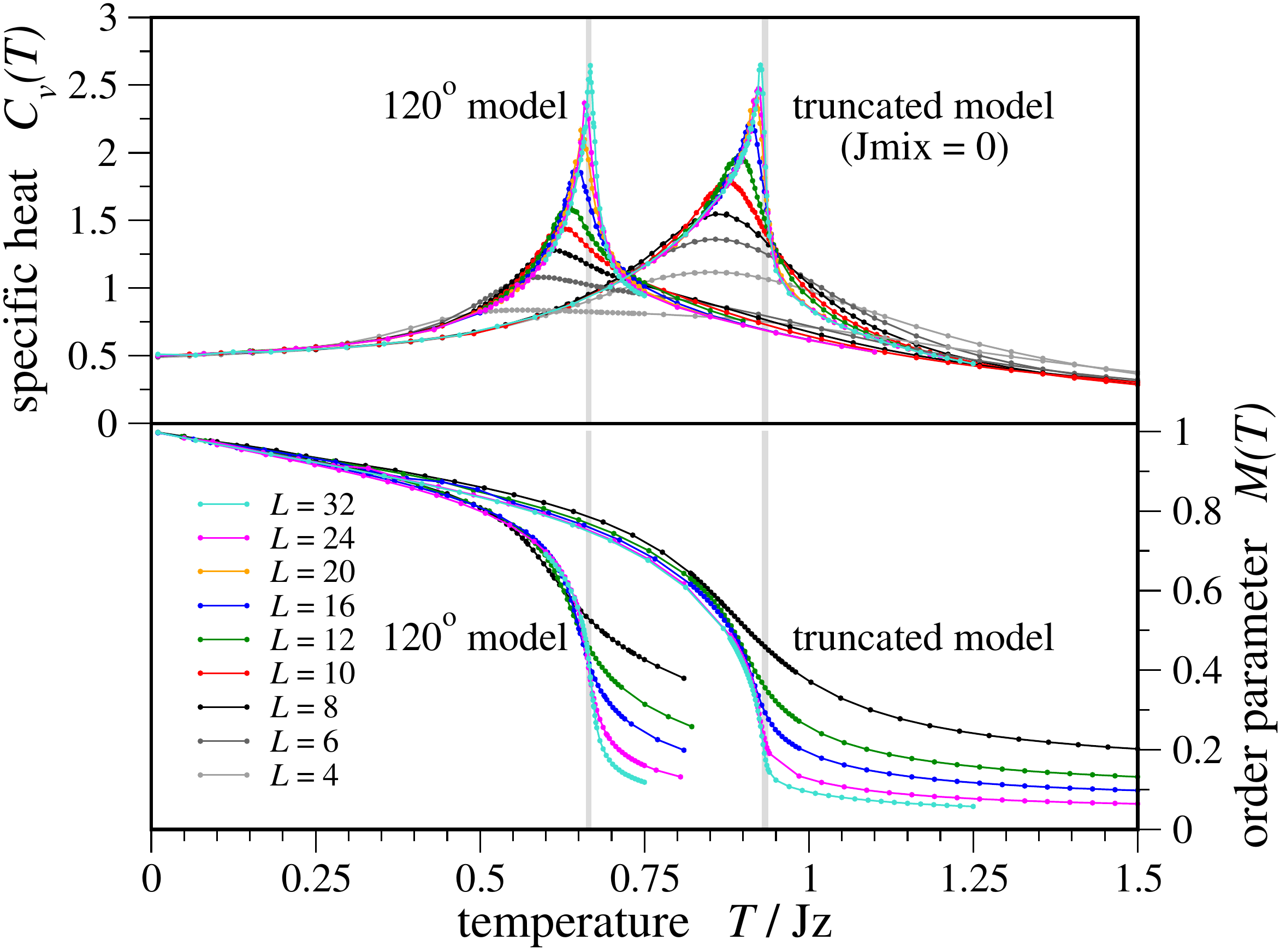}
\end{center}
\caption{
   Orbital ordering transition in the classical model:
   Specific heat $C_v(T)$ (upper panel) and order parameter $M(T)$ (lower panel) 
   versus temperature for various system sizes $L$.  }
\label{Fig:Classical_Cv}
\end{figure}

We now turn to a discussion of the thermodynamic properties of the classical 120\degree\ model, 
in particular the thermal ordering transition into the low-temperature states described above.
To investigate the latter we have run extensive Monte Carlo simulations of model \eqref{eq:H120},
going well beyond previous numerics for the (diluted) symmetric model \cite{Ishihara05}.
Concentrating on a family of models, where we vary $J_{\rm mix} / J_z$ but keep $J_x = J_z$ fixed,
we find a line of {\sl continuous} thermal phase transitions. Fig.~\ref{Fig:Classical_Cv} shows the specific heat
$C_v(T)$ at the transition diverging with linear system size $L$ for two members in this family, the symmetric model 
with $J_{\rm mix} = J_z$ and a `truncated' model where we drop the mixing terms in Hamiltonian \eqref{eq:H120},
i.e.~$J_{\rm mix} = 0$.
For all models in this family we can capture the transition to the low-temperature ordered states by a single
order parameter $M$ (independent of $J_{\rm mix} \leq J_z$). 
Since in our numerical simulations we do not know a priori which one of the three possible ordering planes the system 
spontaneously selects at the ordering transition, we define the order parameter as the maximum of the $xy$, $xz$, and $yz$ 
plane magnetizations
   $M = \text{max}(M_{xy}, M_{xz}, M_{yz})$,
where the magnetization in the $xy$ plane is given by 
$M_{xy} = \sum_{z}\left| \sum_{x,y} \mathbf{T}_{x,y,z} \right|$
and $M_{xz}, M_{yz}$ are obtained by cyclic permutations of the indices in the sums. 
As expected this order parameter quickly grows at the transition temperature $T_c$
 (see the lower panel of Fig.~\ref{Fig:Classical_Cv}).
Despite the relatively large system sizes studied here, finite-size effects still render the
identification of the universality class of these transitions somewhat ambiguous, reminiscent
of studies of similar models in two spatial dimensions \cite{ScrewBoundary}.
Tracking the ordering temperature $T_c$ with the strength of the mixing term $J_{\rm mix}$, as shown in Fig.~\ref{Fig:Tc_Jmix},
we find a significant suppression for the symmetric 120\degree\ model, for which the
transition occurs around $T_c/J_z = 0.677 \pm 0.003$. 

\paragraph{The quantum model.--}

\begin{figure}[b]
\begin{center}
  \includegraphics[width=\columnwidth]{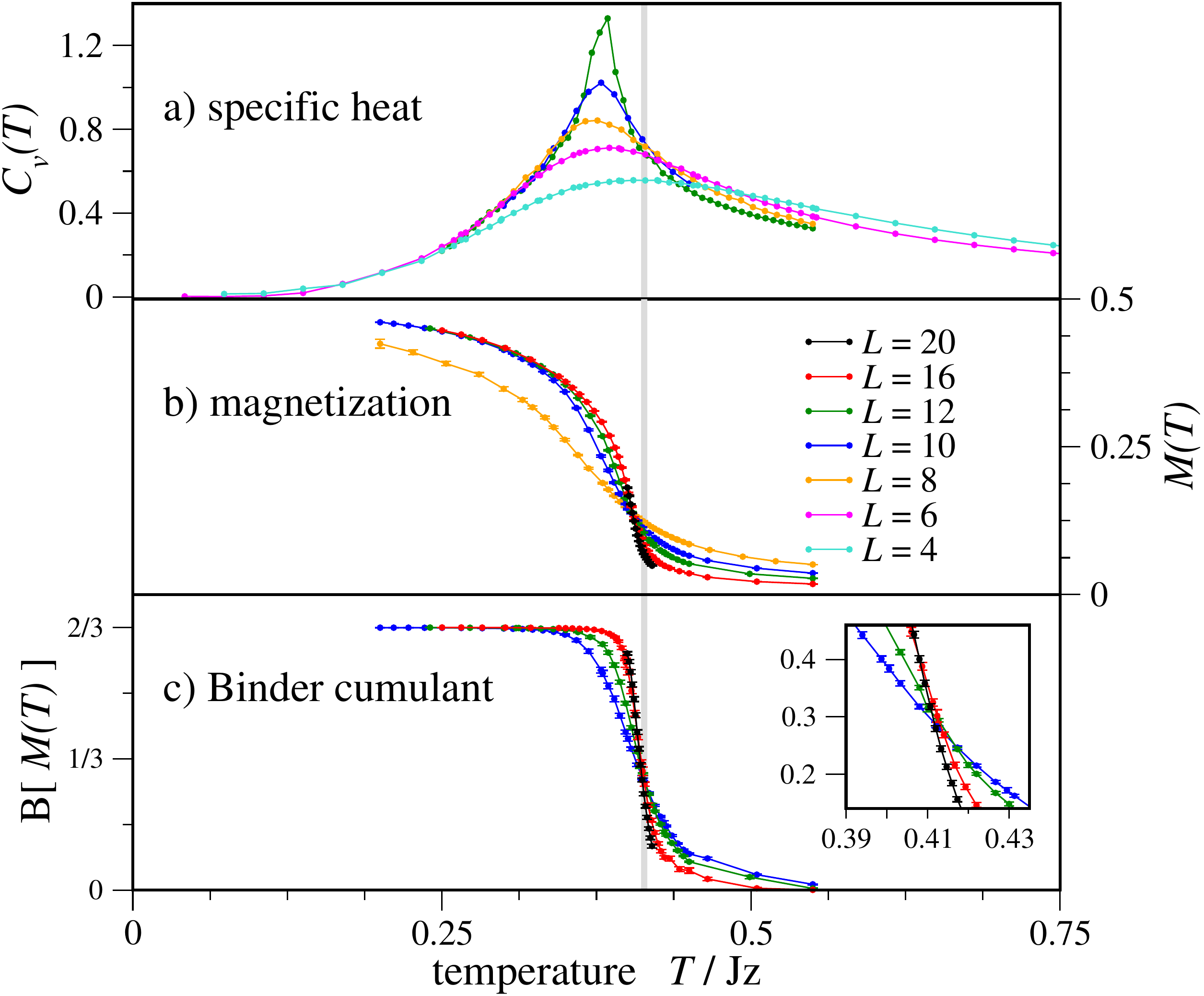}
\end{center}
\caption{
   Orbital ordering transition in the quantum model:
   a) specific heat $C_v(T)$, b) magnetization $M(T) = |\sum_i T^z_i|$,
   and c) Binder cumulant $B\left[ M(T) \right] = 1 - \langle M(T)^4 \rangle / 3 \langle M(T)^2 \rangle^2$.
   The latter exhibits a crossing point for data of different system sizes $L$, strongly indicative of a continuous transition
   \cite{FootnoteQuantumModel}.
 }
\label{Fig:QuantumSpecificHeat}
\end{figure}

As an inroad into exploring ground states and thermodynamics of the quantum 120\degree\ model in an extended $(J_x/J_z, J_{\rm mix}/J_z)$ 
parameter space around the symmetric point, we first concentrate on a family of models for which the mixing term $J_{\rm mix}$ vanishes
but for which we can still vary $J_x/J_z$. This line in parameter space stands out as it allows for a thorough analysis using quantum Monte
Carlo (QMC) simulations, while all other regions in parameter space of non-zero $J_{\rm mix}$ are plagued by the so-called sign problem. 
Along this line, we first discuss the `truncated' model at equal coupling $J_x = J_z$, for which
we have run extensive QMC simulations using an extension of the ALPS looper code~\cite{ALPS,LooperCode}.
Our numerical findings, summarized in Fig.~\ref{Fig:QuantumSpecificHeat}, show that 
this model undergoes a {\sl continuous} thermal phase transition around $T_c/J_z = 0.41 \pm 0.01$ into an orbitally ordered state at low
temperatures. In this ordered state all orbitals are found to spontaneously orient in either the $\ket{3z^2-r^2}$ or $\ket{x^2-y^2}$ orbital configurations,
corresponding to pseudospins pointing in the $\pm T^z$ directions as indicated in Fig.~\ref{Fig:QuantumSpecificHeat}b). 
This ordered orbital state, which we call the `$T^z$ polarized' state, precisely corresponds to the $0\degree$ and $180\degree$ states found
as low-temperature states in the classical truncated model, indicating that thermal fluctuations and quantum effects favor the same states.

\begin{figure}[t]
\begin{center}
  \includegraphics[width=\columnwidth]{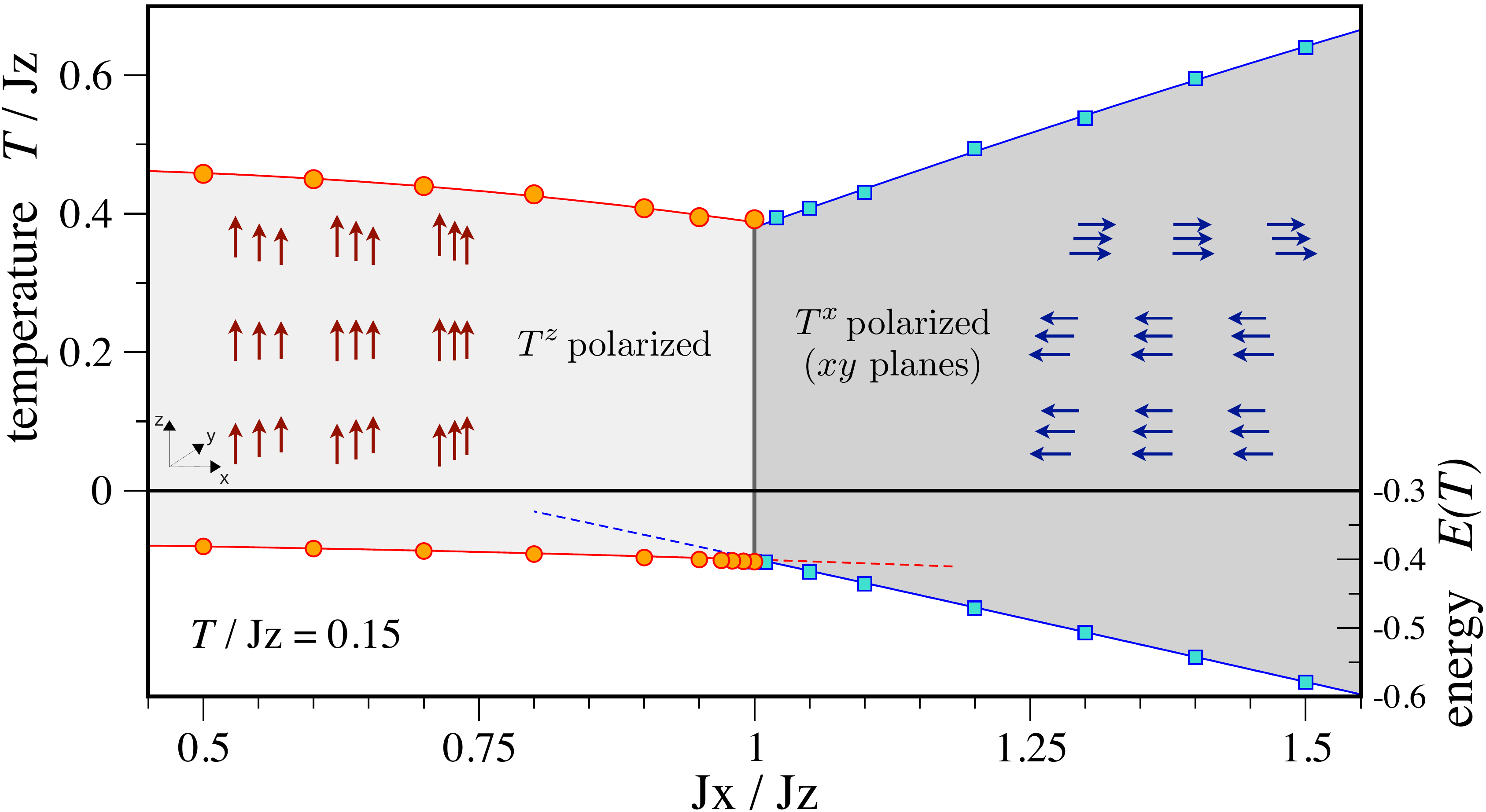}
\end{center}
\caption{
   Variation of the transition temperature (upper panel) and ground-state energy (lower panel)
   for the quantum model.
 }
\label{Fig:QuantumTransitionTemperature}
\end{figure}

\begin{figure}[t]
\begin{center}
  \includegraphics[width=\columnwidth]{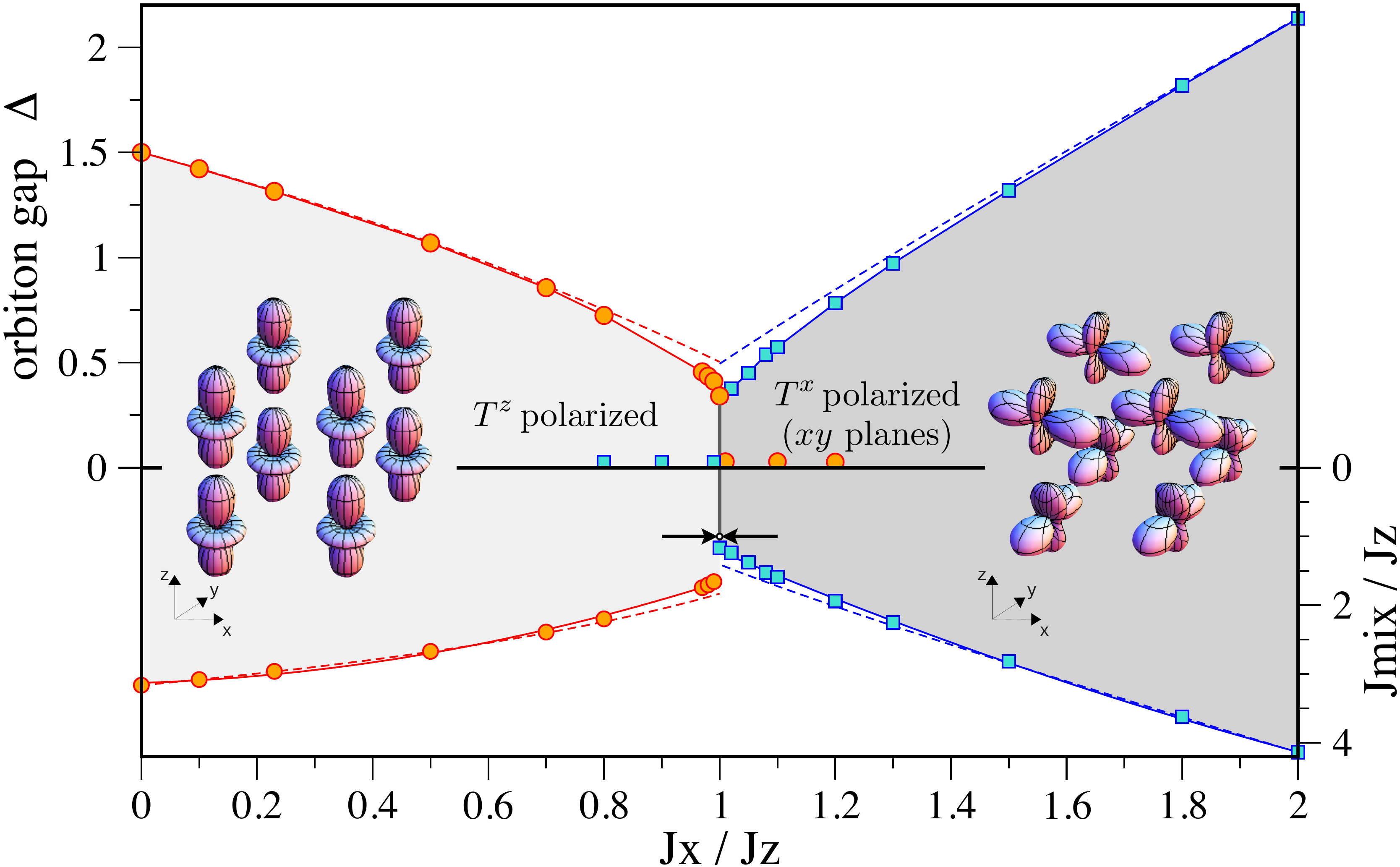}
\end{center}
\caption{
   Phase diagram of the quantum 120\degree\ model:
   The upper panel shows the orbiton gap of the orbital polarized states as a function of $J_x/J_z$ obtained from 
   QMC simulations at $T=0.15$ (symbols)
   and 3rd order perturbation theory (dashed lines).
   The lower panel shows the critical value of $J_{\rm mix}$ at which the orbiton gap closes.
   The location of the symmetric 120\degree\ model is indicated by the arrows. 
 }
\label{Fig:QuantumPhaseDiagram}
\end{figure}

As we vary $J_x/J_z$ away from equal coupling we find that 
the truncated model above exhibits another peculiarity: it sits right at a first-order transition between 
different quantum ground states \cite{FootnoteCompassModel}.
For $J_x < J_z$ this is the same $\pm T^z$ polarized state found for equal coupling (at finite temperature),
while for $J_x > J_z$ the pseudospins in any given $xy$-plane point along the $\pm T^x$ directions 
corresponding to $(\ket{3z^2-r^2} \pm \ket{x^2 - y^2})/\sqrt{2}$ orbital configurations 
(illustrated in the insets of Figs.~\ref{Fig:QuantumTransitionTemperature} and \ref{Fig:QuantumPhaseDiagram}), 
but pseudospins in different $xy$-planes do not have to be aligned.
This first-order transition is apparent in the level crossing of energies shown in the lower 
panel of Fig.~\ref{Fig:QuantumTransitionTemperature} calculated from both QMC simulations (at temperatures well below the thermal transition)
and 2nd order $T=0$ perturbation expansions around the limits of $J_x=0$ and $J_x \to \infty$.
The rather good agreement of the two approaches indicates that quantum effects only modestly change the ground states  with varying $J_x/J_z$.


Both orbitally ordered phases discussed above exhibit gapped elementary excitations
corresponding to a single pseudospin flip, e.g. an `orbital flip' 
$\ket{3z^2 - r^2} \leftrightarrow \ket{x^2 - y^2}$ in the $T^z$ polarized state.
We can directly estimate the excitation gap $\Delta(T)$ of such an `orbiton' excitation in our QMC simulations
\cite{LooperCode,GapCalculation}
as
$
 \Delta(T) = 2\pi\, T  \left( \frac{S(0)}{S(2\pi\, T)} - 1 \right)^{-1/2} ,
$
where $S(\omega)$ is a Fourier transform on an imaginary time correlation function $\sum_r C(r,\tau)$
of the pseudospins.
Our results are given in the top panel of Fig.~\ref{Fig:QuantumPhaseDiagram}. Again we find good quantitative agreement with
3rd order perturbative results (dashed lines) calculated around the limits of $J_x=0,\infty$.
For both orbitally ordered phases, the orbiton gap is suppressed as one approaches the truncated model at equal coupling, 
where for both phases we measure a gap of $\Delta(J_x=J_z) = (0.34 \pm 0.04) J_z$.
As a consequence, the temperature of the thermal phase transition is also significantly suppressed in the vicinity of the truncated model 
(see 
Fig.~\ref{Fig:QuantumTransitionTemperature}).

Having established the nature of the ordered phases for vanishing mixing term, 
we are now in a position to return to a discussion of the 120\degree\ model in the full $(J_x/J_z, J_{\rm mix}/J_z)$ parameter space. 
While a small mixing term $J_{\rm mix} \neq 0$ will not affect these gapped phases, 
a sufficiently large mixing term can close the orbiton gap.
For large $J_{\rm mix}$, a mean-field approximation indicates an `orthogonal' ordered orbital state
where the pseudospins are aligned within each $xz$ (or $yz$) plane, but aligned approximately perpendicular to each other between planes.
Assuming that the instability of the polarized phases indeed arises primarily from orbiton condensation 
(and is not preempted by some other transition), we can map out a phase boundary in parameter space
by calculating a $T=0$ perturbation expansion of the orbiton gap in $J_{\rm mix}/J_z$.
Our results are shown in the lower panel of Fig.~\ref{Fig:QuantumPhaseDiagram}, where the lines indicate the closing of the orbiton gap
when considering 2nd order corrections 
to the gap values for a given ratio $J_x/J_z$ calculated either from QMC
simulations (symbols) or perturbation theory (dashed lines). 
Interestingly, these phase boundaries intersect the $J_x = J_z$ axis at values of $J_{\rm mix}/J_z$ just above one, 
the location of the symmetric 120\degree~model. 
Given that the perturbative results {\sl overestimate} the critical value of $J_{\rm mix}/J_z$, one might be tempted to conclude that
the symmetric model is right at a multicritical point between the three phases in this extended phase diagram. 
In particular, this would indicate that the quantum ground states found for the truncated model at $J_{\rm mix}/J_z = 0$ adiabatically
connect to the ground states at the symmetric point. However, at the symmetric point the rotational symmetry requires that 
if the $\ket{3z^2 - r^2}$ ($\ket{x^2 - y^2}$) orbital states (i.e.~the $\pm T^z$ polarized states) remain ground states, 
then also their symmetry related $\ket{3x^2 - r^2}$ ($\ket{y^2 - z^2}$) and $\ket{3y^2 - r^2}$ ($\ket{z^2 - x^2}$) orbital states 
must become ground states.
This leaves us with two possible scenarios to connect the $T^x$ polarized states of the truncated model
to the symmetric point: 
i) Quantum effects for non-zero $J_{\rm mix}$ have the same effect as thermal fluctuations in the classical model 
   and these states adiabatically turn into a combination of the symmetry required states above (with two more states coming down from higher energies).
   Some support for this scenario comes from considering a 1/S expansion for varying $0 \leq J_{\rm mix}/J_z \leq 1$ (generalizing previous
   calculations \cite{OrbitalDynamics,Kubo}), which in linear order gives a zero-point energy that exactly mimics the behavior of the free energy
   obtained for the classical model in Fig.~\ref{Fig:F_Jmix}, i.e. two minima at 90\degree\ and 270\degree\ found in the vicinity of the truncated model bifurcate with increasing 
   $J_{\rm mix}/J_z$.
ii) The $T^x$ polarized state remains unchanged and brings in four more symmetry related states.
Scenario i) thus selects exactly the same six ordering states at angles $0\degree, 60\degree, \ldots$ as in the classical symmetric model, 
while for scenario ii) we get twelve ordering states at angles $0\degree, 30\degree, 60\degree\, \ldots, 330\degree$.

We acknowledge discussions with L. Balents, A. Laeuchli, Z. Nussinov, J. van den Brink, and M. Zhitomirsky.




\begin{thebibliography}{00}
\bibitem{FootnoteSign}
On the cubic lattice we can transform the antiferromagnetic 120\degree\ model~\cite{120model} 
to a ferromagnetic one through rotations of all pseudospins by an angle $\pi$ about the $T^y$ axis on one
sublattice.
\bibitem{FootnoteQuantumModel}
The continuous nature of this transition is further corroborated by the observation 
of no bimodal histograms of the energy, magnetization or other observables.
\bibitem{FootnoteCompassModel}
This peculiarity of our truncated model is reminiscent of the quantum compass model at equal coupling \cite{KK}, 
which also is situated at a $T=0$ first-order transition \cite{FirstOrderQuantumCompass},
but whose thermodynamics slightly differ in that the compass model undergoes a directional ordering transition \cite{DirectionalOrderingTransition}.
\end{thebibliography}
\end{document}